\documentclass[aps,prd,preprint,superscriptaddress,nofootinbib]{revtex4}
\usepackage{graphics}
\usepackage{epsfig}
\usepackage{latexsym}
\usepackage{colordvi}
\usepackage{amsmath}
\usepackage{amssymb}

\newcommand{\be}{\begin{equation}}
\newcommand{\ee}{\end{equation}}
\newcommand{\bea}{\begin{eqnarray}}
\newcommand{\eea}{\end{eqnarray}}

\def\IB{\relax\hbox{$\inbar\kern-.3em{\rm B}$}}
\def\IC{\relax\hbox{$\inbar\kern-.3em{\rm C}$}}
\def\ID{\relax\hbox{$\inbar\kern-.3em{\rm D}$}}
\def\IE{\relax\hbox{$\inbar\kern-.3em{\rm E}$}}
\def\IF{\relax\hbox{$\inbar\kern-.3em{\rm F}$}}
\def\IG{\relax\hbox{$\inbar\kern-.3em{\rm G}$}}
\def\IGa{\relax\hbox{${\rm I}\kern-.18em\Gamma$}}
\def\IH{\relax{\rm I\kern-.18em H}}
\def\IK{\relax{\rm I\kern-.18em K}}
\def\IL{\relax{\rm I\kern-.18em L}}
\def\IP{\relax{\rm I\kern-.18em P}}
\def\IR{\relax{\rm I\kern-.18em R}}
\def\IZ{\relax{\rm Z\kern-.5em Z}}

\usepackage{amsfonts}

\def\drawbox#1#2{\hrule height#2pt
        \hbox{\vrule width#2pt height#1pt \kern#1pt
              \vrule width#2pt}
              \hrule height#2pt}

\def\Fund#1#2{\vcenter{\vbox{\drawbox{#1}{#2}}}}
\def\Asym#1#2{\vcenter{\vbox{\drawbox{#1}{#2}
              \kern-#2pt       
              \drawbox{#1}{#2}}}}

\def\fund{\Fund{6.4}{0.3}}
\def\asymm{\Asym{6.4}{0.3}}

\begin{document}

\title{A model of light dark matter and dark radiation}

\author{Deog Ki Hong}
\email[E-mail: ]{dkhong@pusan.ac.kr} \affiliation{Department of
Physics,   Pusan National University,
             Busan 46241, Korea}

\date{\today}

\begin{abstract}
We propose a model for dark matter and dark radiation, based on a strongly-coupled dark ${\rm SU}(5)$ gauge theory with fundamental and decuplet dark-quarks. The model supports light dark-baryons,  respecting the chiral symmetry, which are electrically neutral but have electromagnetic form factors, and also a light dark-axion. Since the coupling of dark baryons to the (electrically charged) standard model particles is inversely proportional to the square of the confinement scale, dark baryons become either hot dark matter or cold dark matter, depending on when the dark color confines. For the confinement scale $\Lambda\sim 10-10^3~{\rm GeV}$ the dark baryons of mass about $ 1~{\rm GeV}-1~{\rm MeV}$ become cold dark matter with naturally small magnetic moment and give the correct relic abundance.  
\end{abstract}


\maketitle

\newpage

\section{Introduction}
Dark matter (DM) is widely believed to be one of the major components that make up our universe~\cite{Akrami:2018vks}. Though its origin is uncovered yet, the most popular explanation for DM is that it is made of weakly interacting massive particles (WIMP) that naturally arise with mass of a few hundred GeV in the models beyond the standard model (BSM) of particle physics~\cite{Jungman:1995df}. Such weakly interacting particles  are  thermally  produced in early universe to give the correct relic density of our present universe. But, it has so far escaped all the searches that have been performed until now.  Another popular theoretical candidate for DM is the axion, which is the pseudo Nambu-Goldstone boson of  the spontaneously broken ${\rm U}(1)_{\rm PQ}$ symmetry,  introduced to solve the strong CP problem in the standard model~\cite{Peccei:1977hh}. If the axion decay constant or the scale of the Peccei-Quinn (PQ) symmetry breaking is $f_{\rm PQ}\sim 10^{9}-10^{12}~{\rm GeV}$, then the axion becomes very light,  $m_{\rm axion}\sim 10^{-2}- 10^{-5}~{\rm eV}$, and abundantly produced in early universe to become the significant component of DM at present~\cite{Preskill:1982cy,Abbott:1982af,Dine:1982ah,Kim:2008hd}. 

Recently however there have been lots of activities in theory and also in experiments to probe the dark matter much lighter than a typical WIMP~\cite{Alexander:2016aln}.
Among them the strongly interacting massive particles (SIMP) have been studied intensively. The SIMP models are not directly related to the BSM models that address the problems of the standard model, but they offer new windows for the dark matter searches to probe vast ranges of the DM mass~\cite{Hochberg:2014dra}. Furthermore, having significant self-interactions, they could explain the problems in the structure formation of our universe that WIMP models fail to explain~\cite{Hochberg:2014kqa}. 

In this paper we propose a new model of DM that contains a light and stable dark-baryon doublet, which is neutral but has a small magnetic moment, inversely proportional to the square of the confining scale of dark forces that bind the baryon.
The dark baryons are light because the chiral symmetry protects their mass, thus the model is natural in the sense of 't Hooft~\cite{tHooft:1979rat}. 
If one component of the dark-baryon doublet is almost massless, it becomes dark radiation. The model is based on a strongly coupled gauge theory that is confined  but does not break all the chiral symmetries of the theory in the infrared (IR). In the confined phase the spin-$\frac12$ dark-baryons do exist and saturate the flavor anomalies of the ultraviolet (UV) sector of the theory.  The dark baryons therefore remain massless in the IR without breaking the chiral symmetry~\footnote{Another consistent solution to the 't Hooft anomaly matching would be the broken chiral symmetry in the IR as in quantum chromodynamics (QCD), where the Nambu-Goldstone bosons saturate the anomaly.}  by the 't Hooft anomaly matching condition~\cite{tHooft:1979rat}. 
For the dark-baryon DM,  we break the chiral symmetry explicitly to give the dark baryons small mass and a magnetic moment, keeping them neutral however. The dark baryons therefore couple to the standard model particles electromagnetically via dipole interactions and are produced thermally in the early universe to constitute DM of our universe.  The model has in addition a very light dark-axion or dark $\eta^{\prime}$ that is potentially a good candidate for the hot DM.  

Our model is a minimal, ultraviolet-complete model for the so-called dipolar dark matter~\cite{Sigurdson:2004zp} that provides naturally small magnetic-moments for the dark matter, while electrically neurtal. 


\section{Dark matter Model }

Consider a ${\rm SU}(5)$ gauge theory with two massless fundamental dark-quarks $q_i^a$ ($a=1,2$) and two massless decuplet dark-quarks $Q_{ij}^{\alpha}$ ($\alpha=1,2$) in the second-rank antisymmetric tensor representation, where $i,j=1,\cdots,5$ are the ${\rm SU}(5)$ dark-color indices~\footnote{Our model can be generalized to ${\rm SU}(2n+1)$ for $n=2,3,\cdots$ with a doublet of fundametal dark-quarks and another doublet of dark-quarks in $n$ anti-symmetric index representations.}. Both quarks carry a dark-baryon number ${\rm U}(1)_B$,  the dark-axial charge ${\rm U}(1)_A$, and also the electric charges. (See Table~\ref{matter}.) The theory then has the chiral symmetry $G_f\otimes G_{as}$, where
\begin{equation}
G_f={\rm SU}(2)^f_L\times {\rm SU}(2)^f_R\quad{\rm and}\quad G_{as}={\rm SU}(2)^{as}_L\times {\rm SU}(2)^{as}_R\,.
\end{equation}
\begin{table}[ht]
\caption{The matter content of the light dark-baryon model}
\centering
\begin{tabular}{c| c c  c c c c}
\hline
 & ${\rm SU}(5)$ & ${\rm SU}(2)^f$ & ${\rm SU}(2)^{as}$ & ${\rm U}(1)_B$ & ${\rm U}(1)_A$ & ${\rm U}(1)_{\rm em}$  \\ [0.5ex] 
\hline
$q_i^a$ &\ $\fund$ &\ $\fund$ &\ $1$ &\ $\frac15$ &\ $q_f$ &\ $\frac25$ \\
$Q_{ij}^{\alpha}$ &\ $\asymm$ &\ $1$ &\ $\fund$ &\ $\frac25$ &\ $q_{as}$ &\ $-\frac15$ \\
${\chi}^a$ &\ $1$ &\ $\fund$  &\ $1$ &\ $1$ &\ $q_f+2q_{as}$ &\ $0$ \\ [1ex]
\hline
\end{tabular}
\label{matter}
\end{table}

\subsection{'t Hooft anomaly matching and chiral symmetry breaking}
As shown by 't Hooft~\cite{tHooft:1979rat}, the low-energy spectrum of gauge theories is highly constrained by the flavor anomalies. 
Since the flavor anomalies are invariant under the renormalization group flow, the flavor anomaly of the ultraviolet sector has to match that of the infrared sector, which is nontrivial for the confining theories such as the non-abelian gauge theories with small number of flavors. Indeed, Coleman and Witten showed by using the 't Hooft anomaly matching condition  that the chiral symmetry in QCD with the number of flavors more than two has to be broken~\cite{Coleman:1980mx}. Similarly one can show that the chiral symmetry ${ G}_{as}$ of dark quarks in the antisymmetric representation of the ${\rm SU}(5)$ gauge theory  has to be broken, because otherwise the 't Hooft anomaly matching condition is not satisfied in the infrared. 
However, we find that the chiral symmetry $G_f$ of the fundamental dark-quarks does not have to be broken, since the flavor anomaly of the fundamental dark-quarks is matched in the IR by the massless spin-$\frac12$ baryons. 
\begin{figure}
\centering%
            \includegraphics[width=2.5in]{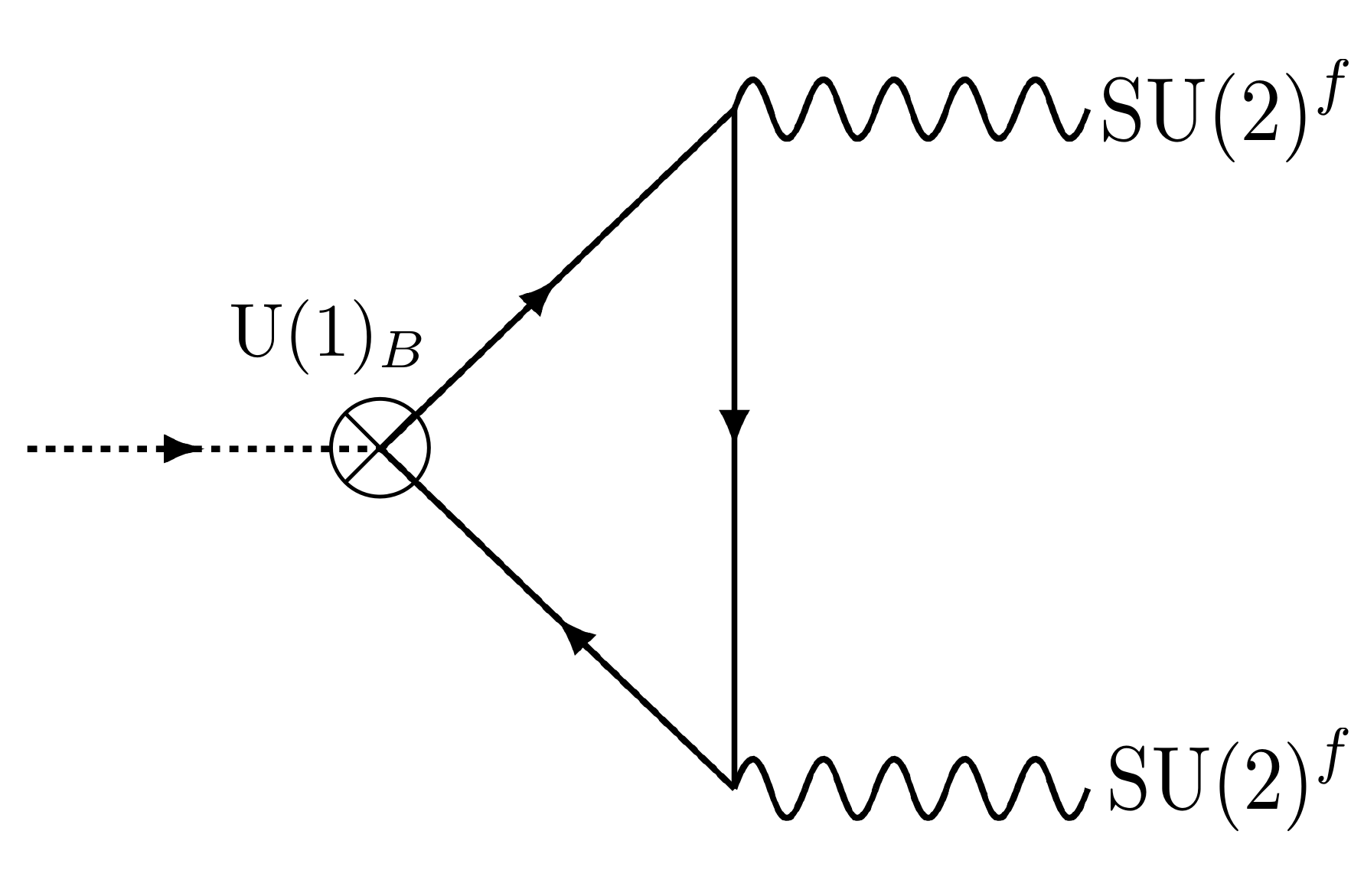}
\caption{The ${\rm U}(1)_B\otimes\left[{\rm SU}(2)^f\right]^2$  flavor anomaly of the ${\rm SU}(5)$ theory. }   
\label{fig1}
\end{figure}

When dark-quarks of both fundamental and second-rank antisymmetric representations are present, the ${\rm  SU}(5)$ gauge theory supports a spin-$\frac12$ dark-baryon, called a chimera dark-baryon, made of dark-quarks in two different representations, given as, suppressing the spin indices, 
\begin{equation}
\chi^{a}\sim \epsilon_{\alpha\beta}q_i^{a}Q^{\alpha}_{jk}Q_{lm}^{\beta}\epsilon^{ijklm}\,,
\end{equation}
where $i,j,k,l,m=1,..,5$ are the color indices and $a,\alpha,\beta=1,2$ are the flavor indices. We note that the chimera dark-baryons are spin-$\frac12$ and form a doublet under the ${\rm SU}(2)^{f}$ flavor symmetry of the fundamental dark-quarks, while being a singlet under the ${\rm SU}(2)^{as}$ flavor symmetry of the dark-quarks in the antisymmetric second-rank representation. 

The triangle anomaly, shown in Fig.~\ref{fig1}, is given as, assuming the chimera dark-baryons are massless,
\begin{equation}
(p+q)^{\lambda}\Gamma^{ab}_{\mu\nu\lambda}=-\frac{1}{\pi^2}A^{ab}\epsilon_{\mu\nu\lambda\sigma}p^{\lambda}q^{\sigma}\,,
\end{equation}
where
\begin{equation}
\Gamma^{ab}_{\mu\nu\lambda}(p,q,-p-q)=\int d^4x\, d^4y\,e^{ip\cdot x}e^{iq\cdot y}\left<Tj^{a}_{\mu}(x)j^{b}_{\nu}(y)j^{B}_{\lambda}(0)\right>
\end{equation}
with the ${\rm SU}(2)$ flavor currents ($T^A=\tau^a$) and the ${\rm U}(1)_B$ dark-baryon currents ($T^A=B$) defined for the UV and IR  theories respectively as
\begin{equation}
j^{A}_{\mu}=\bar q_L\gamma_{\mu}T^A q_L\,~~ ({\rm UV})\,;
\quad
j^{A}_{\mu}=\bar \chi_L\gamma_{\mu}T^A \chi_L\,~~ ({\rm IR})\,.
\end{equation}
We find that the coefficients of the UV and IR anomalies match, 
\begin{equation}
A_{\rm UV}^{ab}=\frac15\cdot 5\,{\rm Tr}\left(\tau^a\tau^b\right),
\quad A_{\rm IR}^{ab}=1\cdot {\rm Tr}\left(\tau^a\tau^b\right)\,,
\end{equation} 
where $\tau^{a}$'s are the ${\rm SU}(2)$ generators.
On the other hand, the flavor anomaly of ${\rm U}(1)_B\otimes\left[{\rm SU}(2)_L^{as}\right]^2$ is not matched in the infrared, because the chimera baryon is singlet under  ${\rm SU}(2)^{as}$, while the antisymmetric second-rank dark-quarks are a doublet.  
Therefore in the IR only the chiral symmetry $G_{as}$ of the two-index antisymmetric dark-quarks $Q_{ij}^{\alpha}$ has to be spontaneously broken down to its vectorial subgroup, leading to three massless dark-pions $\pi^A$ ($A=1,2,3$):
\begin{equation}
G_f\otimes G_{as}\mapsto G_f\otimes{\rm SU}(2)_V^{as}\,.
\end{equation}

\subsection{The chiral effective Lagrangian for light dark baryons and dark mesons}
Since the chimera dark-baryons are singlet under the flavor symmetry ${\rm SU}(2)$ of the decuplet quarks, $Q_{ij}$, they couple to the dark pions or the small fluctuations of $Q^{\alpha}\bar Q_{\beta}$ condensates, only through the flavor-singlet operators of dark pions, which is therefore quite supporessed. The leading interaction terms between them are the following dimension 8 operators:
\begin{equation}
{\cal L}_{\rm int}\ni \frac{c_8}{f_{\pi}^4}\bar \chi\!\not\!\partial\chi\left(\partial_{\mu}\pi^A\right)^2\,,
\end{equation}
where $f_{\pi}$ is the dark-pion decay constant. 

The ${\rm U}(1)_A$ symmetry is in general anomalous and the dark-axion $a$ or the flavor singlet Nambu-Goldstone boson of
the $\left<Q^{\alpha}\bar Q_{\beta}\right>$ condensate couples to the dark ${\rm SU}(5)$ gauge fields with the field strength tensor $G_{\mu\nu}$ and its dual ${\tilde G}_{\mu\nu}$, 
\begin{equation}
{\cal L}_{da}\ni \frac{c_a}{16\pi^2}\frac{a}{f_a}\,{\rm Tr}\,G_{\mu\nu}\tilde G^{\mu\nu}\,,
\end{equation}
where the dark-axion coupling 
\begin{equation}
c_a=q_fC_2(F)+q_{as}C_2(AS)\,.
\end{equation}
By choosing the axial charges of dark-quarks, $q_f=7$ and $q_{as}=-3$, to be inversely proportional to the eigenvalues of the quadratic Casimirs\,\footnote{For ${\rm SU}(N)$, $C_2(F)/C_2(AS)=(N^2-1)/2(N-1)(N+2)$.}, we cancel the axial ${\rm SU}(5)$ anomaly, $c_a=0$, to have the dark-axion lighter than or comparable to dark pions~\cite{Weinberg:1975ui}
\begin{equation}
m_a\lesssim \sqrt{3}\,m_{\pi}\,.
\end{equation}
If we choose the current mass for the decuplet dark-quarks, $m_Q={\rm diag}(m_U,m_D)$, such that $m_U\ll m_D$, the dark axion becomes much lighter than dark pions, $m_a\ll m_{\pi}$. 
The major components of dark matter in our model are then the light dark-baryons and very light dark-axions. They couple
to each other at the leading-order as 
\begin{equation}
{\cal L}_{a\chi}\ni\frac{c_{a\chi}}{f_a}\partial_{\mu}a\,\bar \chi\gamma^{\mu}\gamma_5\chi\,,
\end{equation} 
which gives the $3\to2$ process of $\chi\chi \bar \chi\mapsto \chi a$, that is needed for the SIMP model.

\subsection{Very light dark-axions}
Though the dark ${\rm SU}(5)$ anomaly is absent,  the dark axions decay into two photons by the axial anomaly\,\footnote{The dark axions will decay into two dark baryons, if heavier than twice of dark baryons, $m_a>2m_{\chi}$, which could be seen at Belle II as $e^+e^-\to\gamma+{\not \!\!E_T}$.}, 
\begin{equation}
{\cal L}_{a\gamma\gamma}=\frac{c_{\gamma}}{32\pi^2 }\cdot\frac{6a}{f_a}\,F_{\mu\nu}{\tilde F}^{\mu\nu}\,,
\label{agamma}
\end{equation}
where $F_{\mu\nu}$ and $\tilde {F}_{\mu\nu}$ are the (dual) field-strength tensors of the photon fields and $c_{\gamma}=\left[\left(\frac25\right)^2\cdot 5\cdot2+\left(\frac15\right)^2\cdot 10\cdot 2\right]e^2 =\frac{12}{5}e^2$\,.

If we introduce a current mass $m_Q\ll\Lambda$ for the decuplet dark-quark $Q_{\ij}^{\alpha}$, 
it will lift the vacuum degeneracy to induce a potential for the dark-axion fields, given at the leading order in $m_Q$ as
\begin{equation}
V(a)=m_Q\,\Lambda^3\,\cos\left(\frac{6a}{f_a}\right)\,,
\label{axion_pot}
\end{equation}
where $f_a$ is the dark-axion decay constant and $\Lambda$ is defined by the decuplet dark-quark condensation, 
\begin{equation}
\left< Q_{\alpha}\bar Q_{\beta}\right>=\Lambda^3\delta_{\alpha\beta}\,.
\end{equation}
The condensation scale $\Lambda$ is of the order of the confinement scale below which all the excitations are dark-color singlets\,\footnote{We assume that the scales of confinement and chiral symmetry breaking are same, as the recent lattice calculation for the ${\rm SU}(4)$ gauge theory shows indeed they are equal~\cite{Ayyar:2018ppa}.}.
The dark axion gets a mass $m_a$ that satisfies
\begin{equation}
f_a^2m_a^2\simeq 36m_Q\Lambda^3\,.
\end{equation}
Since the dark-axion decay constant is of the order of the condensation scale  $f_a\sim \Lambda$, the dark-axion mass is
given as  
\begin{equation}
m_a\sim \sqrt{m_Q\,\Lambda}\ll\Lambda\,.
\end{equation}
\section{Relic abundances}
\subsection{Light dark-baryon and cold relics}
In the infrared our ${\rm SU}(5)$ DM model has a doublet of massless Dirac (chimera) dark-baryons that are electrically neutral, though their constituents carry electric charges. Since the chiral symmetry of the dark-quarks in the antisymmetric second-rank representation is spontaneously broken, they get a dynamical mass, though confined. As the 't Hooft anomaly matching indicates, it is however possible to form massless (dark) baryons with massive constituents~\cite{Dimopoulos:1981xc}.  In order for them to be a good candidate for DM, we give them a small (Dirac) mass, $m_{\chi}$, which should be independent of the confining scale $\sim\Lambda$ but proportional to the current mass of the fundamental dark-quark, $m_q$, because the dark-baryon mass vanishes in the chiral limit\,\footnote{Since the current mass, $m_Q$, for the decuplet dark-quark does not break  $G_f$ or the chiral symmetry of the dark-baryon, the dark-baryon mass does not depend on $m_Q$ at least at the leading order in the small mass limit.} or $m_q\to0$~\cite{tHooft:1979rat,Preskill:1981sr}.

When the dark baryon has a small Dirac mass, it breaks the chiral symmetry and will naturally induce a magnetic dipole moment, $\mu_{\chi}$, for the dark baryon, since its constituents are electrically charged:
\begin{equation}
\mu_{\chi}=g\frac{e}{2m_{\chi}}\,,
\end{equation}
where the gyromagnetic ratio $g$ is a non-perturbative quantity and will deviate significantly from 2, the Dirac value, for the dark baryon.  One can estimate its value in certain approximations such as the vector meson dominance or the gauge-gravity duality~\cite{Hong:2007kx,Hong:2007ay,Hong:2007dq}, which generally works well in the confined gauge theory for the large-color limit, $N_c\to\infty$.   

\begin{figure}
\centering%
            \includegraphics[width=5in]{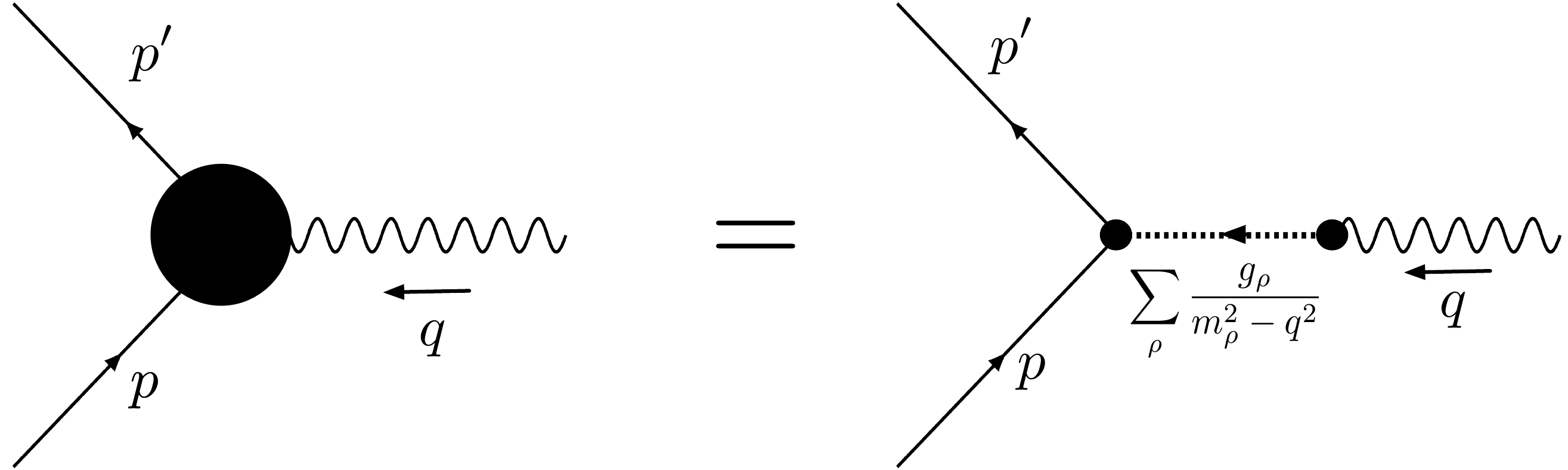}
\caption{Vector dominance in the form factor}   
\label{fig2}
\end{figure}

The matrix elements of an external electromagnetic current for the dark-baryon spinor $u_{\chi}(p)$ can be written as, with $q=p^{\prime}-p$,   
\begin{equation}
\bar u_{\chi}(p^{\prime}) \left[\gamma^{\mu}F_1(q^2)+\frac{i\sigma^{\mu\nu}q_{\nu}}{2m_{\chi}}F_2(q^2)\right]u_{\chi}(p)\,,
\end{equation} 
where the Pauli form factor is given in the gauge-gravity duality (See Fig.~\ref{fig2}) as
\begin{equation}
F_2(-Q^2)=\sum_n\frac{g_{\rho_n}}{m_{\rho_n}^2+Q^2}\,,
\end{equation}
where $\rho_n$ is the $n$-th vector meson with its coupling to the baryon $g_{\rho_n}$\,.
In the confining gauge theory the vector meson mass, $m_{\rho}\sim\Lambda$. The coupling $g_{\rho}$ measures the overlap of the wavefunctions of the vector mesons and the baryons in the holographic direction~\cite{Hong:2007dq}.
The $g$ factor is therefore given by the dimensional analysis as 
\begin{equation}
g\equiv F_2(0)=\kappa\frac{m_{\chi}^2}{\Lambda^2}\,,
\end{equation}
where $\kappa={\cal O}(1)$ is a dimensionless constant.
If the chiral symmetry were spontaneously broken, the baryon mass would be of the order of the confinement scale $m_{\chi}\sim\Lambda$ and the $g$-factor would be of the order one as in the case of nucleons. However, since the chiral symmetry is not spontaneously broken for the chimera dark-baryon, its gyromagnetic ratio will be parametrically very small, $g\sim (m_{\chi}/\Lambda)^2\ll1$\,, as $m_{\chi}\sim m_q\ll\Lambda$\,.

Being electrically neutral but carrying a very small magnetic-dipole-moment, the chimera dark-baryons, which are the lightest fermions in the dark sector, are natural candidates for the dark matter. In fact they belong to the class of the dipolar dark matter (DDM), studied in~\cite{Sigurdson:2004zp}, whose magnetic moment is naturally much smaller though, 
given as
\begin{equation}
\mu_{\chi}=g\cdot\mu_B\left(\frac{m_e}{m_{\chi}}\right)\,,
\end{equation}
where $m_e$ is the electron mass and $\mu_B=1.23\times 10^{-10}{\rm \,e\,cm}$ is the Bohr magneton. 

As the universe expands to undergo the confinement phase transition of  the dark-colors, most of the dark hadrons except those light ones, protected by symmetry, will decouple thermally right after the phase transition. As the universe expands further, the chimera dark-baryons will be out of equilibrium, becoming cold DM, and the total number of the dark baryons freezes out. The freeze-out temperature of the chimera dark-baryon is determined by the condition that the collision rate of dark baryons is equal to the expansion rate of our universe, $\left<n\sigma\,v\right>_{T_f}=H$. 
The freezeout temperature $T_f$ or $x_f\equiv m_{\chi}/T_f$ is approximately~\cite{Kolb:1990vq} 
\begin{equation}
x_f\simeq \ln\left[A/\sqrt{\ln A}\right]
\end{equation}
where $A=0.038\sqrt{g_*} \,m_{pl}\,m_{\chi}\left<\sigma v\right>$ with $m_{pl}=1.22\times 10^{19}\,{\rm GeV}$ and $g_*$ being the effective number of relativistic degrees of freedom at $T_f$. 

\begin{figure}
         \begin{minipage}[b]{5cm}
\centering%
            \includegraphics[width=3.5cm]{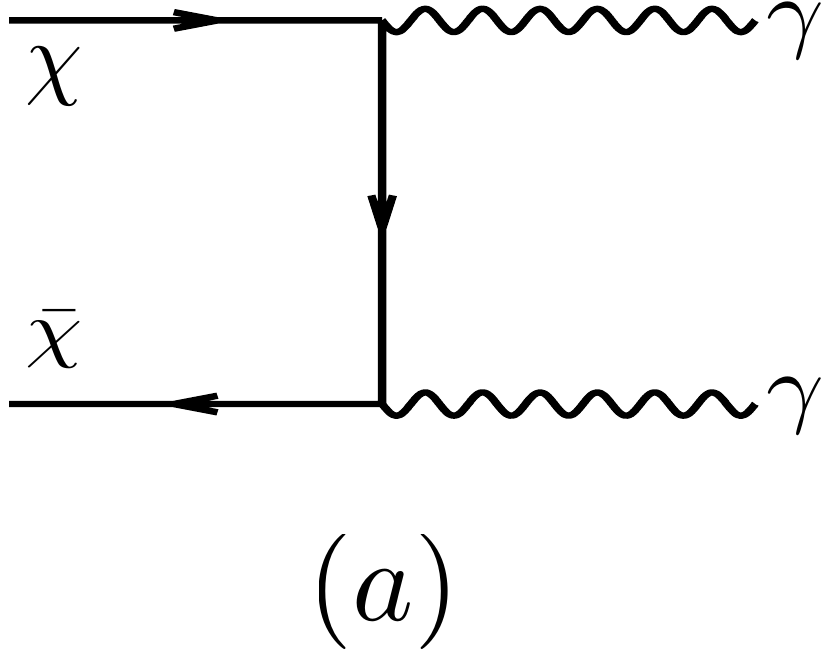}
          \end{minipage}
                  \begin{minipage}[b]{5cm}
\centering%
            \includegraphics[width=3.5cm]{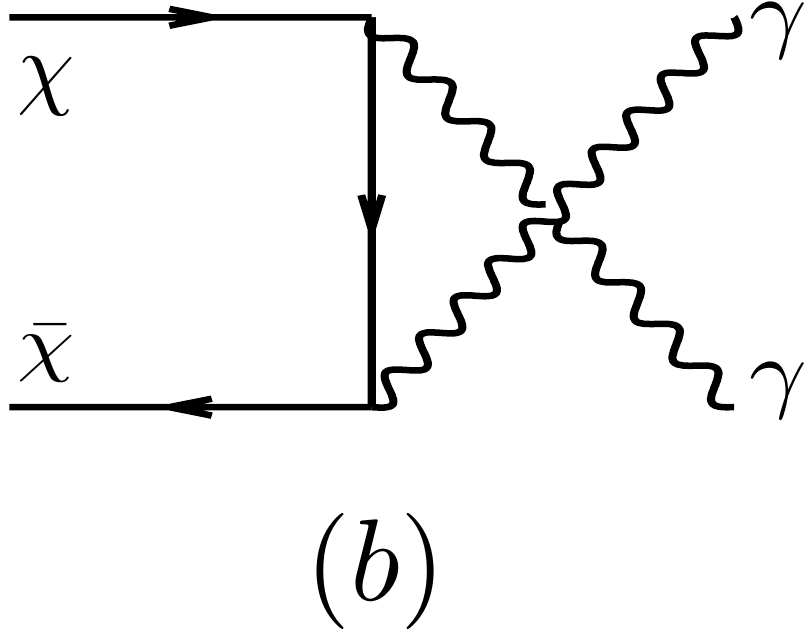}
          \end{minipage}
                  \begin{minipage}[b]{5cm}
\centering%
            \includegraphics[width=3.7cm]{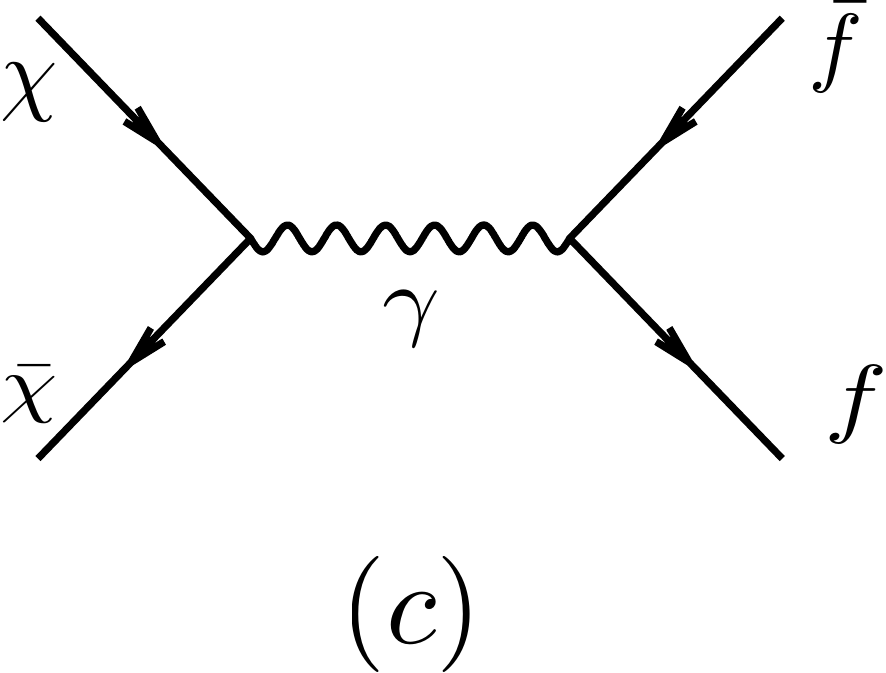}
          \end{minipage}
\caption{Annihilation processes}   
\label{fig3}
\end{figure}

The annihilation processes for the chimera baryons are shown in Fig.~\ref{fig3}.  If $m_{\chi}>m_f$, the annihilation process $(c)$ in Fig.~\ref{fig3} is open and dominant to give the cross section~\cite{Sigurdson:2004zp} 
\begin{equation}
\sigma_{\chi\bar\chi\to f\bar f}v\simeq N_{\rm eff}\,\alpha\,\mu_{\chi}^2\,,
\end{equation}
where $\alpha\simeq1/137$ is the fine-structure constant and $N_{\rm eff}=\sum_fQ_f^2$ is the sum of all electric charge squared in the final states.  
The current relic density of dark baryons, compared to the critical density $\rho_c\equiv 3H_0^2/(8\pi G)$, is given as 
\begin{equation}
\Omega_{\chi}h^2=2.1\times 10^4\left(\frac{m_{\chi}}{m_e}\right)\,\ln\left(A/\sqrt{\ln A}\right)/A\,,
\end{equation}
where $h$ is the current Hubble constant $H_0$ in units of $100~{\rm km\, s^{-1}\, Mpc^{-1}}$.
For $m_{\chi}\sim 1~{\rm MeV}$, $A\simeq4\times 10^5$ or the gyromagnetic ratio $g\simeq10^{-12}$ gives the correct relic density of $\Omega_{\chi}h^2=0.12$~\cite{Akrami:2018vks}, which requires the confinement scale $\Lambda\sim 1~{\rm TeV}$ for the dark ${\rm SU}(5)$. If we increase the mass of the dark baryon, the $g$ factor has to increase for the correct relic abundance. For the benchmark value $m_{\chi}=1~{\rm GeV}$ in~\cite{Sigurdson:2004zp} one finds $g\sim10^{-2}$ for the correct DM relic density and the confinement scale $\Lambda\sim 10~{\rm GeV}$. We therefore find that for $m_{\chi}\sim {\rm a~ few}~{\rm GeV}$ the explicit chiral symmetry breaking by $m_q$ is no longer perturbative and all the low-lying dark-hadrons will have mass of the order of $\Lambda\sim 3-5~{\rm GeV}$ if the chimera dark-baryons  were to be the major components of DM. Our model will be then a kind of SIMP model with dark baryons and dark axion.

\subsection{Hot relics}
\label{hdm}
\begin{figure}
\centering%
            \includegraphics[width=1.6in]{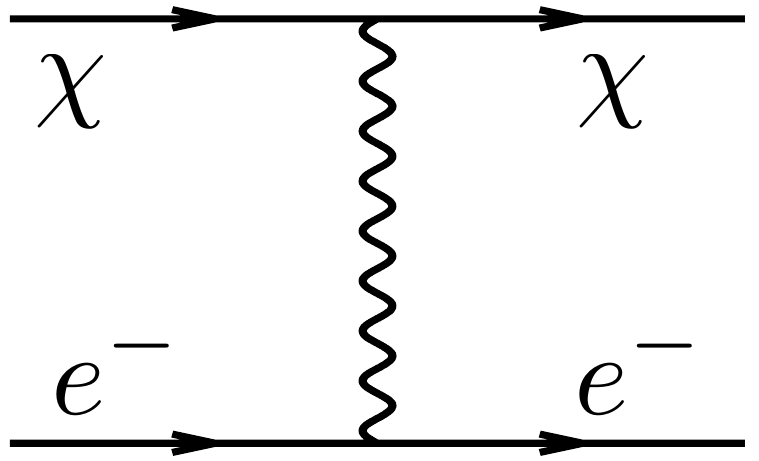}
\caption{A thermal equilibrium process for $m_{\chi}<m_e$.}   
\label{fig4}
\end{figure}

If the dark-baryon is very light, it will decouple from the thermal bath at temperatures much higher than its mass to become hot relics.  The decoupling temperature for the relativistic dark-baryons, whose number density $n(T)=8T^3\zeta(3)/\pi^2$,  is then given as 
\begin{equation}
T_d=\frac{1.67 g_*^{1/2}}{n(T_d)\,T_d^{-3}\cdot \sigma v \cdot m_{pl}} 
\simeq\frac{0.55\,g_*^{1/2}}{\alpha^2{\tilde N}_{\rm eff}}\cdot \frac{\Lambda^4}{m^2_{\chi}m_{pl}}\,,
\label{decoupling}
\end{equation}
where $g_*$ is the effective massless degrees of freedom at $T_d$ and  $\tilde N_{\rm eff}=\sum_fQ_f^2$ is to sum over the electric charge squared for fermions in thermal equilibrium with the dark baryons via the reactions shown in Fig.~\ref{fig4}.  In order not to spoil the Cosmic Microwave Background (CMB) measurements, we require the dark baryons decouple before the  QCD phase transition epoch. The dark-baryon mass is then from Eq.~(\ref{decoupling}) given in terms of  the confinement scale and the decoupling temperature as 
\begin{equation}
m_{\chi}\simeq100\,{\rm eV} \left(\frac{\Lambda}{1\,{\rm GeV}}\right)^2\cdot\left(\frac{1~{\rm GeV}}{T_d}\right)^{1/2}\,.
\end{equation}
The mass of the relativistic dark-baryons, that decouple before the QCD hadronization, therefore should be  $m_{\chi}\gtrsim 1\,{\rm eV}$, since the dark-color has to confine before the decoupling of dark baryons. 

The relativistic degrees of freedom at the decoupling will contribute to the present relic mass density~\cite{Kolb:1990vq}
\begin{equation}
\Omega_{\chi}h^2=7.83\times10^{-2}\left[g_{*}(x_d)/g_{\rm eff}\right]\left(m_{\chi}/{\rm eV}\right)\,.
\end{equation}
If the decoupling temperature $T_d\sim 200~{\rm MeV}$, $g_*(x_d)\sim20$. From the current bound for the relic density $\Omega_mh^2=0.12$~\cite{Akrami:2018vks} and because $g_{\rm eff}=7$ for the dark-baryons, the allowed mass for the hot dark-baryons to be the subdominant component of DM, $\Omega_{\chi}h^2\sim 0.02$, is $m_{\chi}\sim 1~{\rm eV}$, which implies the confinement scale of the dark-color, $\Lambda\sim \Lambda_{\rm QCD}$.  The dark baryons decouple shortly after the dark-color confinement phase transition but just before the QCD hadronization transition.


\subsection{Dark radiation}
When the dark-baryon mass vanishes or the current dark-quark mass $m_q\to0$, the magnetic moment of dark baryons does vanish because of the exact and unbroken chiral symmetry. Though the magnetic moments are absent, the dark baryons do interact with the standard model particles (electrically charged), since the dark-baryons are made of electrically charged dark-quarks. Namely the dark-baryons do have the Dirac form-factor $F_1(-Q^2)$. The leading interaction of dark-baryons to photons is a chiral-invariant dimension 6 operator, 
\begin{equation}
{\cal L}_6=\frac{e\,c_d}{\Lambda^2}\bar\chi\gamma_{\mu}\chi\partial_{\nu}F^{\mu\nu}\,,
\end{equation}
where $e$ is the electromagnetic coupling and $c_d$ is a dimensionless coupling of order one and $\Lambda$ is the confinement scale of the dark-colors.
If we integrate out the photon fields, the scattering amplitude $\chi e^-\to\chi e^-$ in Fig.~\ref{fig4} becomes a four-Fermi operactor, given as 
\begin{equation}
{\cal L}_{4F}=\frac{e^2c_d}{\Lambda^2}\bar\chi\gamma^{\mu}\chi\,\bar \psi_e\gamma_{\mu} \psi_e\,,
\label{dirac}
\end{equation}
where $\chi$ and $\psi_e$ are the dark-baryon fields and the electron fields, respectively. 
Similarly to neutrinos the ratio of the interaction rate to the expansion rate~\cite{Kolb:1990vq} is 
\begin{equation}
\frac{\Gamma_{int}}{H}\sim \frac{e^4c_d^2T^5/\Lambda^2}{T^2/m_{pl}}=\left(\frac{T}{T_{\chi}}\right)^3\,,
\end{equation}
where the decoupling temperature of massless dark-baryons 
\begin{equation}
T_{\chi}\simeq 0.06~{\rm GeV}\,\left(\frac{\sqrt{c_d}\,\Lambda}{1~{\rm GeV}}\right)^{4/3}\,.
\end{equation}
Below $T_{\chi}$ the interaction rate is less than the expansion rate and the massless dark-baryons decouple from the plasma, freely expanding. The massless dark-baryons become dark-radiation, contributing the energy density of the total radiation~\cite{DiValentino:2013qma}, 
\begin{equation}
\rho_{\rm rad}=\left[1+\frac{7}{8}\left(\frac{4}{11}\right)^{4/3}N_{\rm eff}\right]\rho_{\gamma}\,,
\end{equation}
where $\rho_{\gamma}$ is the CMB energy density. The standard model gives $N_{\rm eff}=3.046$ and any extra radiation-like particle will contribute to its deviation, $\Delta N_{\rm eff}$. The contribution from the massless dark-baryons in our model is given as~\cite{Blennow:2012de}
\begin{equation}
\Delta N_{\rm eff}=\frac{13.56}{g_*^s(T_{\chi})^{4/3}}\cdot g_d\,,
\end{equation}
where $g_*^s$ is the effective number of entropy degrees of freedom for a given temperature and the number of degrees freedom  for the dark-baryon doublet $g_d=8$. As long as the decoupling temperature $T_{\chi}$ is higher than the QCD phase transition or the dark-color confinement scale $\Lambda\gtrsim 3~{\rm GeV}$, the massless dark-baryons give $\Delta N_{\rm eff} \lesssim 0.23$, which is consistent with the recent Planck 2018 data of $\Delta N_{\rm eff}<0.30$ at 95\,\%~\cite{Aghanim:2018eyx}.

Finally we note that if the current dark-quark mass were flavor-asymmetric and only one of the current mass vanishes such as $m_q={\rm diag}(0,m)$, then the down dark-baryon $\chi_d$ becomes a hot dark matter of mass about $1~{\rm eV}$, discussed in the section~\ref{hdm}, while the up dark-baryon $\chi_u$ becomes the dark radiation that gives $\Delta N_{\rm eff} \lesssim 0.12$\, because the light degrees of freedom is reduced by half. Both the dark-baryon $\chi_d$ and the dark-radiation $\chi_u$ are stable because the up and down fermion (dark-quark) numbers are separately conserved.

\subsection{Dark axions as hot dark matter}
Since the chiral symmetry $G_{as}={\rm SU}(2)^{as}_L\times {\rm SU}(2)^{as}_R$ of the decuplet dark-quark $Q_{ij}^{\alpha}$ is spontaneously broken, there will be three dark-pions, which are the small fluctuations of the condensate $\left<Q_{\alpha}\bar Q_{\beta}\right>$. Furthermore, since our axial-charge assignment of dark quarks is to cancel the axial ${\rm SU}(5)$ anomaly, there is a flavor-singlet Nambu-Goldstone boson, which we call dark-axion. The Nambu-Goldstone bosons will get small mass, once we introduce a small current mass
$m_Q={\rm diag}\,(m_U, m_D)$ for the dark-quark decuplet, $Q_{ij}^{\alpha}$ that breaks the chiral symmetry explicitly. 
If we take $m_U\ll m_D$, the dark-axion mass is much smaller than the dark-pion mass
\begin{equation}
m_a=\frac{f_{\pi}m_{\pi}}{f_a/6}\cdot\frac{\sqrt{m_Um_D}}{m_U+m_D}\,\ll m_{\pi}\,,
\end{equation}
where the dark-axion and dark-pion decay constants are of the order of the condensation scale, $f_a\sim f_{\pi}\sim \Lambda$. 

For the dark axion to be DM, it should be very light and long-lived. The dominant decay channel  for the light dark-axion is  $a\to \gamma\gamma$ by the axial anomaly, Eq.~(\ref{agamma}). The decay rate is then
\begin{equation}
\Gamma_{a\to\gamma\gamma}=\frac{g_{a\gamma}^2m_a^3}{64\pi}\simeq 5\times 10^{-22}\,{\rm s}^{-1}\left(\frac{m_a}{10^{-3}~{\rm eV}}\right)^3\cdot\left(\frac{1~{\rm TeV}}{f_a}\right)^2 \,,
\end{equation}
where $g_{a\gamma}=216\alpha/5\pi f_a$\,. Since the dark-axion mass is controlled by $m_U$, the dark-axion can be long-lived by taking $m_U\ll m_D\,(\ll\Lambda)$. For instance, if the dark-axion decay constant $f_a=1~{\rm TeV}$, the dark-axions with $m_a<1.6\times 10^{-2}~{\rm eV}$ live longer than the age of the universe.

\begin{figure}
\centering%
            \includegraphics[width=1.5in]{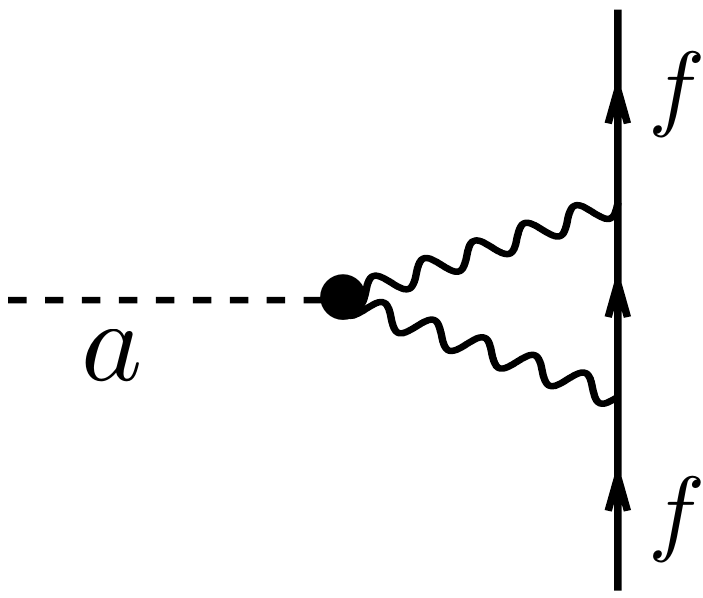}
\caption{Dark-axion $a$ couplings to SM matter fields $f$. The curly line denotes the photons.}   
\label{fig5}
\end{figure}

The dark axions do not directly couple to the standard model (charged) fermions, $f$ at the tree level but only  at the loop level with an effective dark-axion coupling, $c_{af}\sim\alpha^2/f_a$~\cite{Kaplan:1985dv,Srednicki:1985xd} (See Fig.~\ref{fig5}).  From the stellar cooling constraints on the axions~\cite{Raffelt:1994ry} therefore 
\begin{equation}
f_a\sim \Lambda>3\times 10^5~{\rm GeV}\,,
\end{equation}
which turns out to be more stringent than the bound from the dark-radiation constraint.  

\begin{figure}
         \begin{minipage}[b]{6cm}
\centering%
            \includegraphics[width=1.5in]{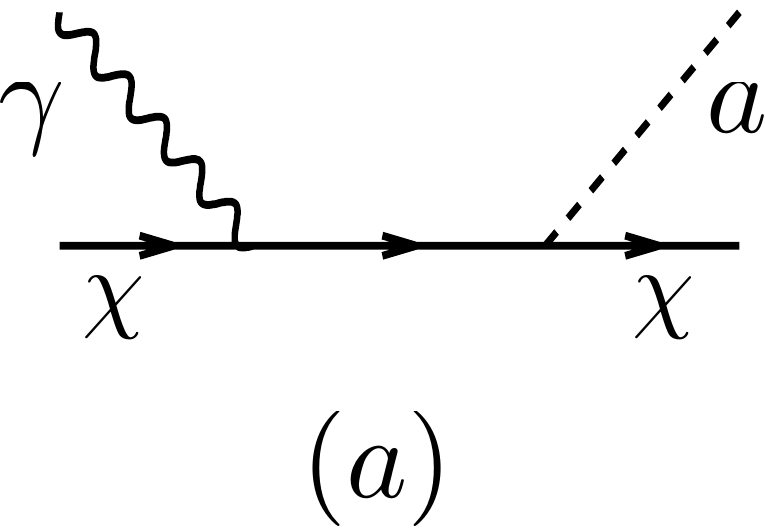}
         \end{minipage}
         \begin{minipage}[b]{6cm}
\centering%
         \includegraphics[width=1.5in]{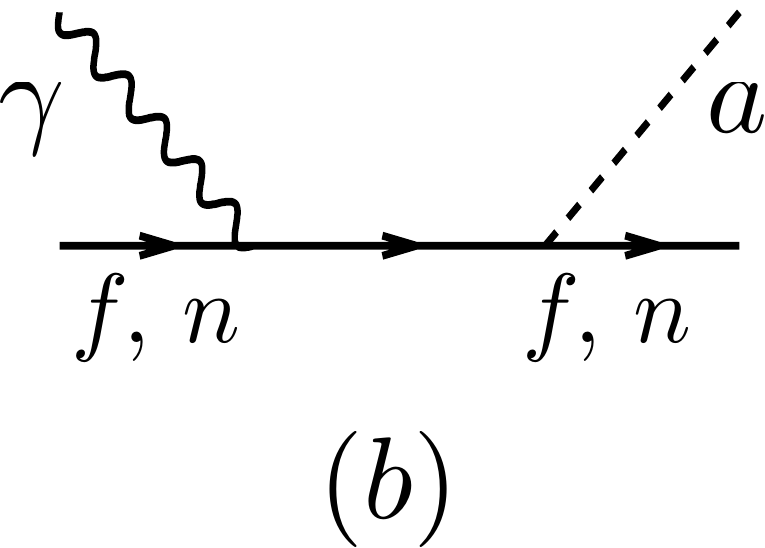} 
        \end{minipage}
\caption{(a) dark-axion thermal process via dark-baryons, $\chi\gamma\leftrightarrow \chi a$. (b) dark-axion scattering with SM fermions, $f$ or nucleons, $n$: $\gamma\,f(n)\leftrightarrow a\, f (n)$. }
\label{fig6}
\end{figure}

In our model the dark axions are produced when the decuplet dark-quarks condense at  temperature $T\sim\Lambda$. 
They will interact with the particles in the thermal plasma by the processes in Fig.~\ref{fig6} and will decouple eventually as the universe expands. Since the scattering cross-section of dark axions is much larger with the standard model fermions or nucleons (Fig.~\ref{fig6}(b)), $\sigma_{\gamma f\to af}\sim \alpha^3/\Lambda^2\gg \sigma_{\chi\gamma\to\chi a}\sim \alpha\, m_{\chi}^4/\Lambda^6$,  dark-axions are dominantly produced by the scattering with quarks or nucleons, similar to the hadronic axions. Therefore, as long as the dark-axions decouple after the QCD transition, $T_f\sim{\Lambda^2}/{m_{pl}\alpha^3}<100~{\rm MeV}$ or $\Lambda>10^6~{\rm GeV}$, the relic abundance of (hot) dark-axions is given as~\cite{Kolb:1990vq} 
\begin{equation}
\Omega_ah^2\simeq \frac{m_a}{130~{\rm eV}}\left(\frac{10}{g_{*S}(T_f)}\right)\,,
\end{equation}
where $g_{*S}$ is the number of entropic degrees of freedom. 
The very light axions with $\Lambda>10^7~{\rm GeV}$ will live long enough to contribute too significantly to the present energy density of the universe, unless $m_a\lesssim1~{\rm eV}$.
If the confining scale of dark-colors is lower than the stellar bound, the dark-axion has to be much heavier than $1~{\rm keV}$ to avoid the astrophysical constraints or  $m_a\gtrsim 1~{\rm MeV}$ to decay well before the primordial nucleosynthesis. In this case the dark-axions are too short-lived to contribute to the relic density, but could be detected at heavy-ion collisions~\cite{Knapen:2016moh}.

Finally we note that the dark-axion becomes the QCD axion in the DFSZ model~\cite{Dine:1981rt,Zhitnitsky:1980tq} that solves the strong CP problem, if we identify the ${\rm U}(1)_A$ as the Peccei-Quinn symmetry and take $m_U$ vanishingly small. The electroweak-singlet PQ field of the DFSZ model is then composite\footnote{In order for the dark axion to couple to gluons, the composite PQ field should couple to two Higgs doublets $\varphi_u$ and $\varphi_d$ of the DFSZ model~\cite{Dine:1981rt,Zhitnitsky:1980tq}. The Higgs doublets therefore have to be composite as well, made of additional dark-quarks such as $\vartheta_{ijk}^a$, fully antisymmetric ${\rm SU}(5)$ indices $i,j,k$.}, the bound state of decuplet dark-quarks, $\varphi_{\rm PQ}\sim \bar Q Q$ and the confining scale of the dark-colors should be the PQ scale, $\Lambda\sim f_{\rm PQ}\sim 10^9-10^{12}~{\rm GeV}$. For such a large confinement scale the magnetic moment of the dark baryons become too small to contribute to the relic density, unless the direct-detection bound is violated~\cite{Sigurdson:2004zp}\,\footnote{For such a small magnetic moment, the leading interaction of the dark baryons will be via the Dirac form factor of the dark baryons rather than the magnetic moment, as in the case of dark radiations~(\ref{dirac}). The dark baryons will couple to the standard model fermions by the four-Fermi interaction but with the coupling $10^{-16}$ smaller than the weak interaction. It is however possible to produce the relic abundance of dark baryons by the so-called freezing-in mechanism through $f{\bar f} \to \chi{\bar \chi}$ annihilation process~\cite{Hall:2009bx,Bernal:2017kxu}.}.

\section{Conclusion and Discussion}
We have proposed a minimal model for light fermionic dark matter of mass between $1~{\rm MeV}-1~{\rm GeV}$, which is neutral but has a (naturally) small magnetic moment with the gyromagnetic ratio $g=10^{-2}-10^{-12}$.  The fermionic dark matter is light because its mass is protected by chiral symmetry. The light dark matter is thermally produced in the early universe to become a cold dark matter and to constitute the current relic abundance. 

The model is based on the confining ${\rm SU}(5)$ gauge theory with a doublet of fundamental dark-quarks and another doublet of decuplet dark-quarks. 
The model has three parameters: The confining scale $\Lambda$ of dark ${\rm SU}(5)$ and the 
current mass, $m_q$  and $m_Q$ for the fundamental and decuplet dark quarks, respectively. Depending on the parameters we have discussed four different scenarios. Among them we find only one scenario provides a viable model for cold dark matter that explains the correct relic abundance of DM. In this model for the correct relic abundance of thermal cold dark matter the confining scale of the dark ${\rm SU}(5)$ has to be $\Lambda= 10-10^3~{\rm GeV}$  and the mass of the lightest dark-baryon is determined to be $1~{\rm MeV}-1~{\rm GeV}$. All other scenarios lead to hot dark matter or dark radiation, which are subdominant components of  DM in our universe. Our model can be used therefore, if not for the thermal cold DM, for a model for dark radiation or (subdominant) hot DM.

Finally we note the dark-axion of our model becomes the QCD axion of the DFSZ model, solving the strong CP problem, if we identify the electroweak-singlet PQ field as the composite of decuplet dark-quarks, provided that the Higgs doublets $\varphi_u$ and $\varphi_d$ of the DFSZ model are also composite of additional dark-quarks and the confining scale $\Lambda$ is as high as the allowed PQ scale, $\Lambda\sim 10^9-10^{12}~{\rm GeV}$. 



\acknowledgments
I would like to thank Ki Young Choi, Anson Hook, Du Hwan Kim, Jihn E. Kim, Tae Hyun Jung, Matt Reece, Josh Ruderman and Tommi Tenkanen for useful conversations. The author is especially grateful to Kwang Sik Jeong and Chang Sub Shin for valuable discussions and also
to the CERN Theory group for the hospitality during his participation at the CERN-Korea TH institute, where this work was initiated. 
This work was supported by a 2-Year Research Grant of Pusan National University.


\begin{thebibliography}{99}

\bibitem{Akrami:2018vks} 
  Y.~Akrami {\it et al.} [Planck Collaboration],
  arXiv:1807.06205 [astro-ph.CO].

\bibitem{Jungman:1995df}
See for a review G.~Jungman, M.~Kamionkowski and K.~Griest,
Phys. Rept. \textbf{267}, 195-373 (1996)
doi:10.1016/0370-1573(95)00058-5
[arXiv:hep-ph/9506380 [hep-ph]].

\bibitem{Peccei:1977hh} 
  R.~D.~Peccei and H.~R.~Quinn,
  Phys.\ Rev.\ Lett.\  {\bf 38}, 1440 (1977).

\bibitem{Preskill:1982cy} 
  J.~Preskill, M.~B.~Wise and F.~Wilczek,
  Phys.\ Lett.\ B {\bf 120}, 127 (1983)
  [Phys.\ Lett.\  {\bf 120B}, 127 (1983)].

\bibitem{Abbott:1982af} 
  L.~F.~Abbott and P.~Sikivie,
  Phys.\ Lett.\ B {\bf 120}, 133 (1983)
  [Phys.\ Lett.\  {\bf 120B}, 133 (1983)].

\bibitem{Dine:1982ah} 
  M.~Dine and W.~Fischler,
  Phys.\ Lett.\ B {\bf 120}, 137 (1983)
  [Phys.\ Lett.\  {\bf 120B}, 137 (1983)].

\bibitem{Kim:2008hd} 
  J.~E.~Kim and G.~Carosi,
  Rev.\ Mod.\ Phys.\  {\bf 82}, 557 (2010)
  [arXiv:0807.3125 [hep-ph]].

\bibitem{Alexander:2016aln} 
 See for instance J.~Alexander {\it et al.},
  arXiv:1608.08632 [hep-ph].

\bibitem{Hochberg:2014dra}
  Y.~Hochberg, E.~Kuflik, T.~Volansky and J.~G.~Wacker,
  Phys.\ Rev.\ Lett.\  {\bf 113} (2014) 171301
  [arXiv:1402.5143 [hep-ph]].

\bibitem{Hochberg:2014kqa}
  Y.~Hochberg, E.~Kuflik, H.~Murayama, T.~Volansky and J.~G.~Wacker,
  Phys.\ Rev.\ Lett.\  {\bf 115} (2015) no.2,  021301
  [arXiv:1411.3727 [hep-ph]].

\bibitem{tHooft:1979rat} 
  G.~'t Hooft,
  NATO Sci.\ Ser.\ B {\bf 59}, 135 (1980).

\bibitem{Sigurdson:2004zp} 
  K.~Sigurdson, M.~Doran, A.~Kurylov, R.~R.~Caldwell and M.~Kamionkowski,
  Phys.\ Rev.\ D {\bf 70}, 083501 (2004)
  Erratum: [Phys.\ Rev.\ D {\bf 73}, 089903 (2006)]
  [astro-ph/0406355].



\bibitem{Coleman:1980mx} 
  S.~R.~Coleman and E.~Witten,
  Phys.\ Rev.\ Lett.\  {\bf 45}, 100 (1980).

\bibitem{Weinberg:1975ui} 
  S.~Weinberg,
  Phys.\ Rev.\ D {\bf 11}, 3583 (1975).

\bibitem{Dimopoulos:1981xc} 
  S.~Dimopoulos and J.~Preskill,
  Nucl.\ Phys.\ B {\bf 199}, 206 (1982).


\bibitem{Preskill:1981sr} 
  J.~Preskill and S.~Weinberg,
  Phys.\ Rev.\ D {\bf 24}, 1059 (1981).

\bibitem{Ayyar:2018ppa} 
  V.~Ayyar, T.~DeGrand, D.~C.~Hackett, W.~I.~Jay, E.~T.~Neil, Y.~Shamir and B.~Svetitsky,
  Phys.\ Rev.\ D {\bf 97}, no. 11, 114502 (2018)
  [arXiv:1802.09644 [hep-lat]].
  
\bibitem{Hong:2007kx} 
  D.~K.~Hong, M.~Rho, H.~U.~Yee and P.~Yi,
  Phys.\ Rev.\ D {\bf 76}, 061901 (2007)
  [hep-th/0701276 [HEP-TH]].
 
\bibitem{Hong:2007ay} 
  D.~K.~Hong, M.~Rho, H.~U.~Yee and P.~Yi,
  JHEP {\bf 0709}, 063 (2007)
  [arXiv:0705.2632 [hep-th]].

\bibitem{Hong:2007dq} 
  D.~K.~Hong, M.~Rho, H.~U.~Yee and P.~Yi,
  Phys.\ Rev.\ D {\bf 77}, 014030 (2008)
  [arXiv:0710.4615 [hep-ph]].
  
    
\bibitem{Kolb:1990vq} 
  E.~W.~Kolb and M.~S.~Turner,
  Front.\ Phys.\  {\bf 69}, 1 (1990).

\bibitem{DiValentino:2013qma} 
  E.~Di Valentino, A.~Melchiorri and O.~Mena,
  JCAP {\bf 1311}, 018 (2013)
  [arXiv:1304.5981 [astro-ph.CO]].

\bibitem{Blennow:2012de} 
  M.~Blennow, E.~Fernandez-Martinez, O.~Mena, J.~Redondo and P.~Serra,
  JCAP {\bf 1207}, 022 (2012)
  [arXiv:1203.5803 [hep-ph]].

\bibitem{Aghanim:2018eyx} 
  N.~Aghanim {\it et al.} [Planck Collaboration],
  arXiv:1807.06209 [astro-ph.CO].

\bibitem{Kaplan:1985dv} 
  D.~B.~Kaplan,
  Nucl.\ Phys.\ B {\bf 260}, 215 (1985).
\bibitem{Srednicki:1985xd} 
  M.~Srednicki,
  Nucl.\ Phys.\ B {\bf 260}, 689 (1985).

\bibitem{Raffelt:1994ry} 
  G.~Raffelt and A.~Weiss,
  Phys.\ Rev.\ D {\bf 51}, 1495 (1995)
  [hep-ph/9410205].
  
\bibitem{Knapen:2016moh} 
  S.~Knapen, T.~Lin, H.~K.~Lou and T.~Melia,
  Phys.\ Rev.\ Lett.\  {\bf 118}, no. 17, 171801 (2017)
  [arXiv:1607.06083 [hep-ph]].

\bibitem{Dine:1981rt} 
  M.~Dine, W.~Fischler and M.~Srednicki,
  Phys.\ Lett.\  {\bf 104B}, 199 (1981).
  
\bibitem{Zhitnitsky:1980tq} 
  A.~R.~Zhitnitsky,
  Sov.\ J.\ Nucl.\ Phys.\  {\bf 31}, 260 (1980)
  [Yad.\ Fiz.\  {\bf 31}, 497 (1980)].

\bibitem{Hall:2009bx} 
  L.~J.~Hall, K.~Jedamzik, J.~March-Russell and S.~M.~West,
  JHEP {\bf 1003}, 080 (2010)
  [arXiv:0911.1120 [hep-ph]].

\bibitem{Bernal:2017kxu} 
  N.~Bernal, M.~Heikinheimo, T.~Tenkanen, K.~Tuominen and V.~Vaskonen,
  Int.\ J.\ Mod.\ Phys.\ A {\bf 32}, no. 27, 1730023 (2017)
  [arXiv:1706.07442 [hep-ph]].
  
\end{thebibliography}
\end{document}